\relax
\documentclass[letterpaper]{article} 
\usepackage{aaai22}  
\usepackage{times}  
\usepackage{helvet}  
\usepackage{courier}  
\usepackage[hyphens]{url}  
\usepackage{graphicx} 
\usepackage{natbib}  
\usepackage{caption} 
\DeclareCaptionStyle{ruled}{labelfont=normalfont,labelsep=colon,strut=off} 
\usepackage{algorithm}
\usepackage{algorithmic}
\usepackage{comment}
\usepackage{subfigure}
\usepackage{amsmath,amssymb}
\DeclareMathOperator{\E}{\mathbb{E}}
\usepackage{multirow} 
\usepackage{newfloat}
\usepackage{listings}
\floatstyle{ruled}
\newfloat{listing}{tb}{lst}{}
\floatname{listing}{Listing}

\title{Improved Sensor-Based Animal Behavior Classification Performance through Conditional Generative Adversarial Network}
\author{
   Zhuqing Zhao,  Dong Ha, Abhishek Damle, Barbara Roqueto dos, Robin White, and Sook Shin Ha\textsuperscript{\rm *}
}
\affiliations{
    Virginia Polytechnic Institute and State University
  
}

\begin{document}

\maketitle

\begin{abstract}
Many activity classifications segments data into fixed window size for feature extraction and classification. However, animal behaviors have various durations that do not match the predetermined window size. The dense labeling and dense prediction methods address this limitation by predicting labels for every point. Thus, by tracing the starting and ending points, we could know the time location, and duration of all occurring activities. Still, the dense prediction could be noisy with misalignments problems. We modified the U-Net and Conditional Generative Adversarial Network (cGAN) with customized loss functions as a training strategy to reduce fragmentation and other misalignments. In cGAN, the discriminator and generator trained against each other like an adversarial competition. The generator produces dense predictions. The discriminator works as a high-level consistency check, in our case, pushing the generator to predict activities with reasonable duration. The model trained with cGAN shows better or comparable performance in the cow, pig, and UCI HAPT dataset. The cGAN-trained modified U-Net improved from 92.17\% to 94.66\% for the UCI HAPT dataset and from 90.85\% to 93.18\% for pig data compared to previous dense prediction work.
\end{abstract}

\section{Introduction}
Farm animals are increasingly managed as large groups to maximize animal production and profit. Nonetheless,  the large groups make monitoring animal health and detecting early signs of disease challenging. Technologies that enable automatic, continuous, and real-time animal behavior monitoring emerged as an alternative solution and received considerable attention. Sensor-based behavior analysis could detect and handle animals’ abnormal behaviors or disease symptoms in the early stage. 

In sensor-based activity recognition, the commonly used approach is to segment motion data into fixed window sizes and extract features for classification. However, this approach could be insufficient: the daily activities of the animals are complex and versatile; different behaviors have various durations that do not match the predetermined window size. 
\begin{figure}[H]
\centering
\includegraphics[width=0.9\columnwidth]{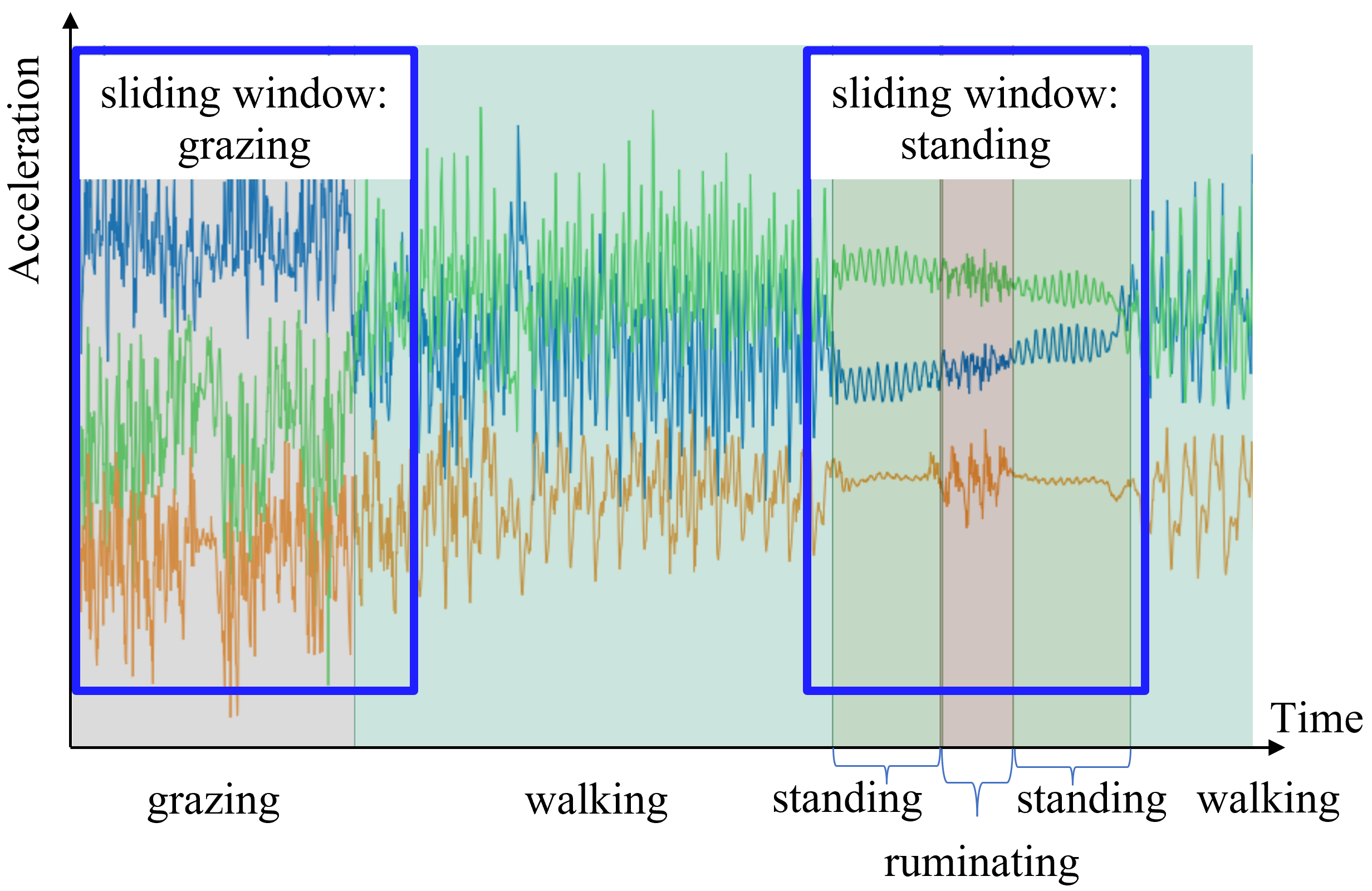} 
\caption{Sliding window labeling problems using cow data as examples: sliding windows labels grazing and suppress the starting part of walking activity; sliding windows labels standing and ignore the ruminating.}
\label{Segmentation problem}
\end{figure}
In figure \ref{Segmentation problem}, the motion data could contain information from multiple behaviors in the sliding window but is labeled with the most frequent activity, where minor activities with shorter durations are ignored. Thus, minimizing the mixing activities and extracting representative information would be key for animal-related behavior classification. Instead of testing different window sizes, we analyzed the dense labeling and prediction methods. We used conditional Generative Adversarial Networks (cGAN) as training strategies. Dense labeling could deliver the precise duration of each activity, especially for transient and scarce activities. The model trained with cGAN could reduce the fragmentation error and achieve better or matching performance. More importantly, the resulted dense prediction maps provide time locations that could be applied for automatic labeling.   

The contributions of this paper are: we are the first to apply dense labeling and prediction in animal activity classification; we are the first to use cGAN to improve model performance in general activity classification to our knowledge; we customized the loss function and modified the U-Net and cGAN for better or comparable performance than the previous state-of-art works.

\section{Related Work}
There are several approaches to animal behavior classification. Many recent studies used image-based classification because visual data usually contains more information compared to sensor data. Some worked with the classification of videos and experimented with different variations to simplify computation \cite{FUENTES2020105627,WU2021106016, LI2022106889,CHEN2020105166}. 

However, visual-related classifications have large amounts of data to process. Besides, it may require identifying and tracking individual animals within a herd to get the accumulated daily activities. An alternative is sensor data-based activity classification because the identity can be found through the sensor attached to each. Many existing works for sensor-based cattle and swing behavior/activity classifications used 
ML algorithms including decision tree  \cite{cowdecisiontree,GONZALEZ201591,ARCIDIACONO2017124}, 
multiple ensemble methods \cite{DUTTA201518}, 
Support Vector Machine (SVM) \cite{MARTISKAINEN200932}, and
a set of the binary classifiers \cite{SMITH201640}, etc. 

In recent years, deep learning gains more attention and models such as Multilayer Perceptron \cite{HOSSEININOORBIN2021106241}, Convolutional Neural Network (CNN) \cite{Pavlovic21124050,Li2021}, and CNN combined with Long Short-Term Memory (LSTM) \cite{PENG2019247} has been explored. The continuous sensor data are segmented into fixed window sizes. 

For each window size, the samples have one label based on the most frequent activity.
These could cause problems when many existing approaches use a large, fixed-size window \cite{MARTISKAINEN200932,GONZALEZ201591,ARABLOUEI2021106045,RAHMAN2018124,cowdecisiontree,BENAISSA2019425}. 
Different activities have various durations that do not match the predetermined window size. Furthermore, many studies show classification performance can be sensitive to the time window size \cite{ROBERT200980,CHANG2022106595}. The segmented data could contain information from multiple activities but labeled with the major activity, whereas activity with a shorter duration is suppressed.

To overcome this shortage, \cite{FCN2017} predicted dense labels for Human Activity Recognition using Fully Convolutional Networks (FCN). The idea was inspired by semantic segmentation to predict labels for every data point. Therefore, the predicted activity map could precisely indicate the time locations of different activities in a single window. However, this method could be insufficient for activities with longer duration owing to problems such as misalignments of activity boundary, substitution, and fragmentation. To improve the previous work, \cite{UNet2019} customized the U-Net model from \cite{unet2015} and used post-correction for fragmentation error. Different from the above method, we modify the U-Net architecture and used cGAN to train the U-Net to reduce the misalignments, so no post-corrections are needed. By reducing the substitution and fragmentation, the dense prediction could better handle activity with a long duration.

The cGAN model was used to train generative model for application such as image tagging \cite{cGAN}. In \cite{pix2pix2017}, cGAN was modified and applied to image to image translation tasks. Many of the U-net and cGAN related work are in visual related tasks \cite{cGANexample1,cGANexample2,cGANexample3}.

\section{Dense Prediction}

\begin{figure}[H]
\centering
\includegraphics[width=0.9\columnwidth]{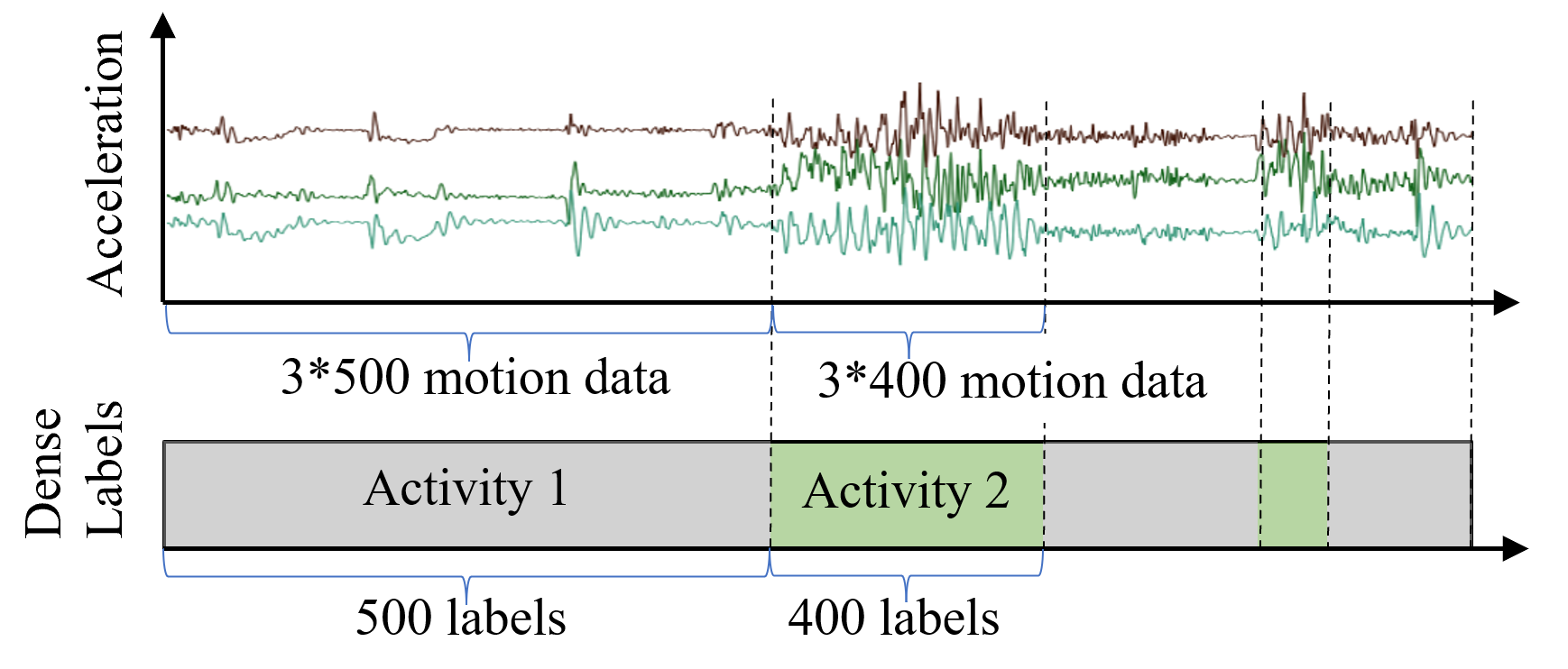}
\caption{Illustration for Dense Labeling.}
\label{dense labeling example}
\end{figure}

Inspired by semantic segmentation to produce pixel-level predictions, our model predicts labels for each data point. The acquired dense predictions are confidence maps that could indicate the location and duration of different activities. The U-Net will learn the general activity patterns of the input sequence and predict the dense labels with matching lengths. For example, in Figure \ref{dense labeling example} the 3-axis acceleration data with 3*500 data points belong to activity 1 and 3*400 labels belong to activity 2. The corresponding dense labels will have the same dimension: 500 points with activity 1 labels and 400 points with activity 2 labels. By tracing the starting and ending points on the dense labels, we could get the relative time location and duration of each activity in the segment.

\subsection{Types of Misalignments}

\begin{figure}[!htb]
\centering
\includegraphics[width=\columnwidth]{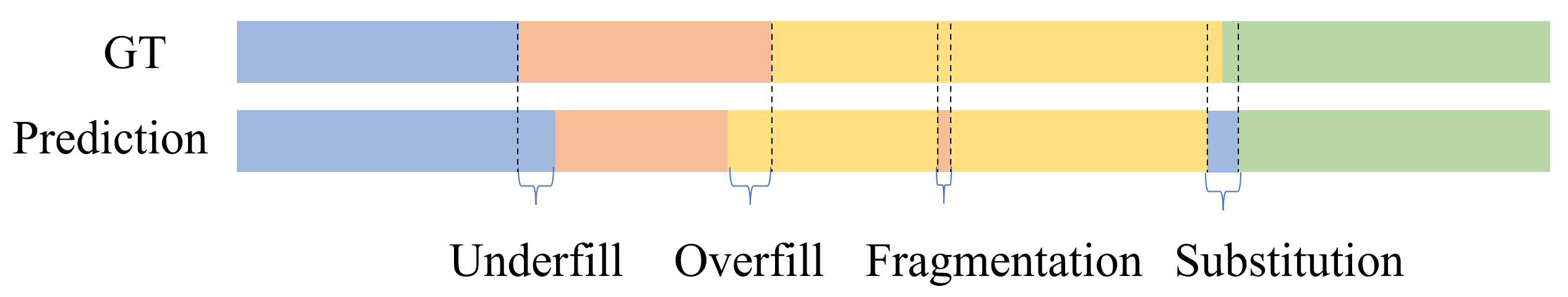} 
\caption{Types of misalignments}
\label{Types of misalignment}
\end{figure}

The general challenges are distinguishing between similar activities and determining the activity boundaries. There are four types of errors: Fragmentation, Substitution, Underfill, and Overfill. The definition is from \cite{UNet2019} and illustrated in Figure \ref{Types of misalignment}.  
\begin{figure*}[ht]
\centering
\includegraphics[width=\textwidth]{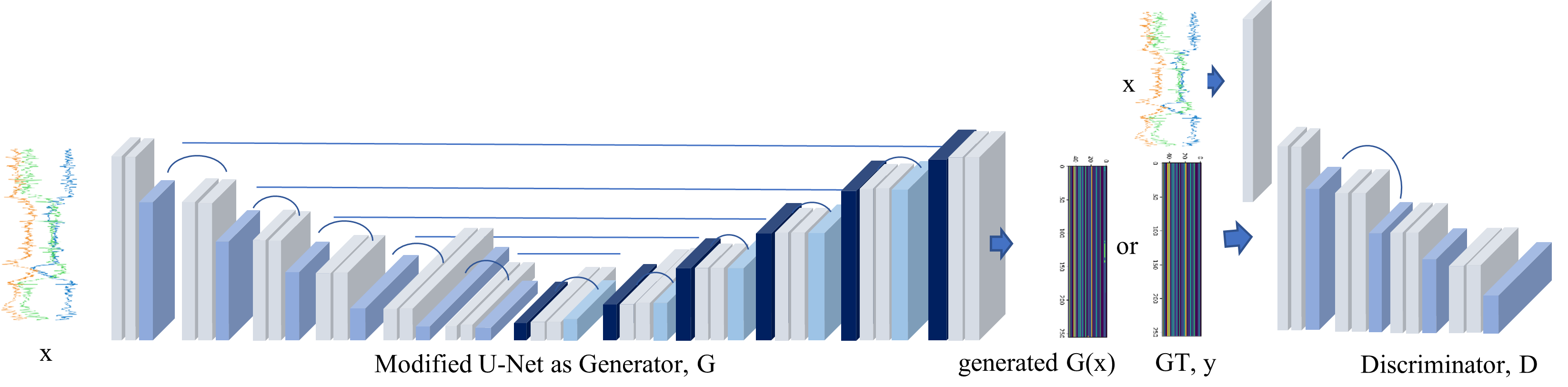} 
\caption{Proposed U-Net trained with CGAN approach: modified U-net as generator takes motion data $x$ to predict dense labels $G(x)$ , and the discriminator$D$ to distinguish between generated dense labels $G(x)$ or ground truth, $y$.}
\label{general idea}
\end{figure*}  
Different activities may not always have distinct features that follow the general pattern. Many activities may have inter-similarity that are difficult to extract unique features to distinguish between others \cite{HARreview2020}. Additionally, activities with shorter duration are more challenging because of the volatile characteristics and the scarcity of the training instance. These problems could confuse the models and cause Fragmentation and Substitution errors in the dense prediction results. Another challenge is the misalignments of activities boundaries: Overfill and Underfill. In time serious data, some activities lack distinct boundaries and cause noisy labels in the labeling process, the models trained with indecisive labels are prone to errors.

    \subsection{Modified U-Net Trained With cGAN}

The general training architecture is similar in \cite{pix2pix2017}. The modified U-Net is the generator for dense prediction. Our designed discriminator is to determine whether the input motion data and densely labeled pairs are ground truth or dense predictions, see Figure \ref{general idea}. This training strategy is to create an adversarial competition. The generator is trained against an adversarial discriminator until the discriminator could not detect the difference between generated dense labels and ground truth. The feedback from the discriminator serves as a consistency check and is reflected in the cGAN loss. As a result, the cGAN loss function pushes the generator to predict label maps that maintain the same characteristic as the ground truth so that the discriminator could not correctly distinguish between the two.
Thus, the objective for cGAN is to minimize the loss function in Equation \ref{lcGAN}. (We did not use random noise vector z for generator input. )
        \begin{equation} \label{lcGAN}
\begin{split}
\mathcal{L}_{cGAN}(G,D)&=\E_{x,y}[logD(x,y)]+\\
             &\E_{x}[log(1-D(x,G(x)))]
\end{split}
\end{equation}
   
The input motion data is the condition and the cGAN structure could be an alternative to the post-correction step targeting fragmentation and substitution errors. The main advantage over post-correction is no additional processes after prediction, so the model is end-to-end trainable. Moreover, the feedback from the discriminator could enforce a high level of consistency between the ground truth and the dense prediction, in our case, pushing the generator to predict activities with reasonable duration and fewer switching activities.

    \subsubsection{Modified U-Net as Generator}
     \begin{figure}[H]
\centering
 \begin{subfigure}[Contraction block.]
		{\includegraphics[width=0.33\columnwidth]{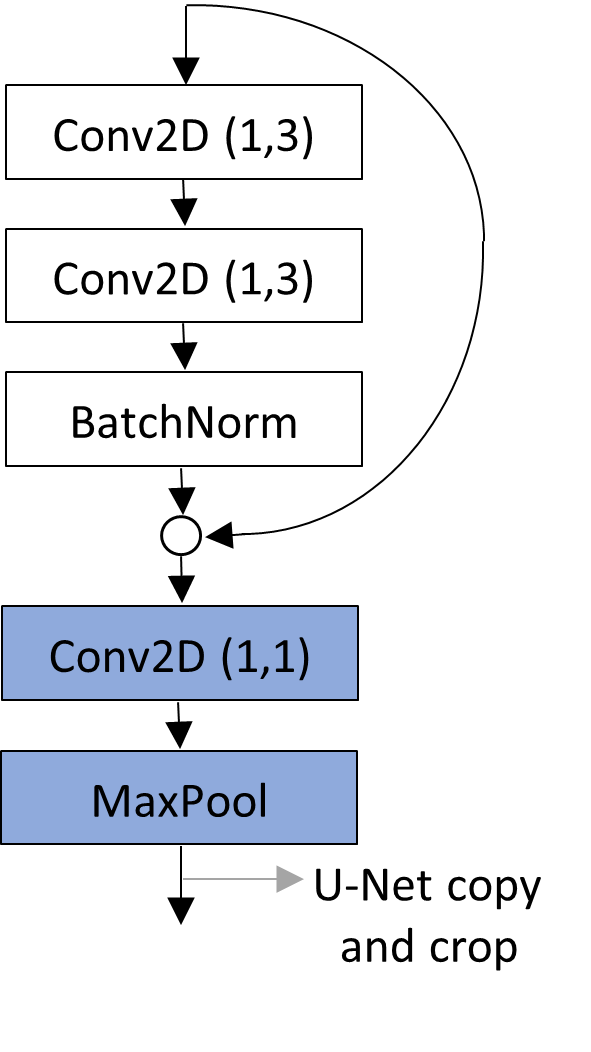}

            \label{contraction}}
		\end{subfigure}
	\begin{subfigure}[Expansion block.]
	{\includegraphics[width=.4\columnwidth]{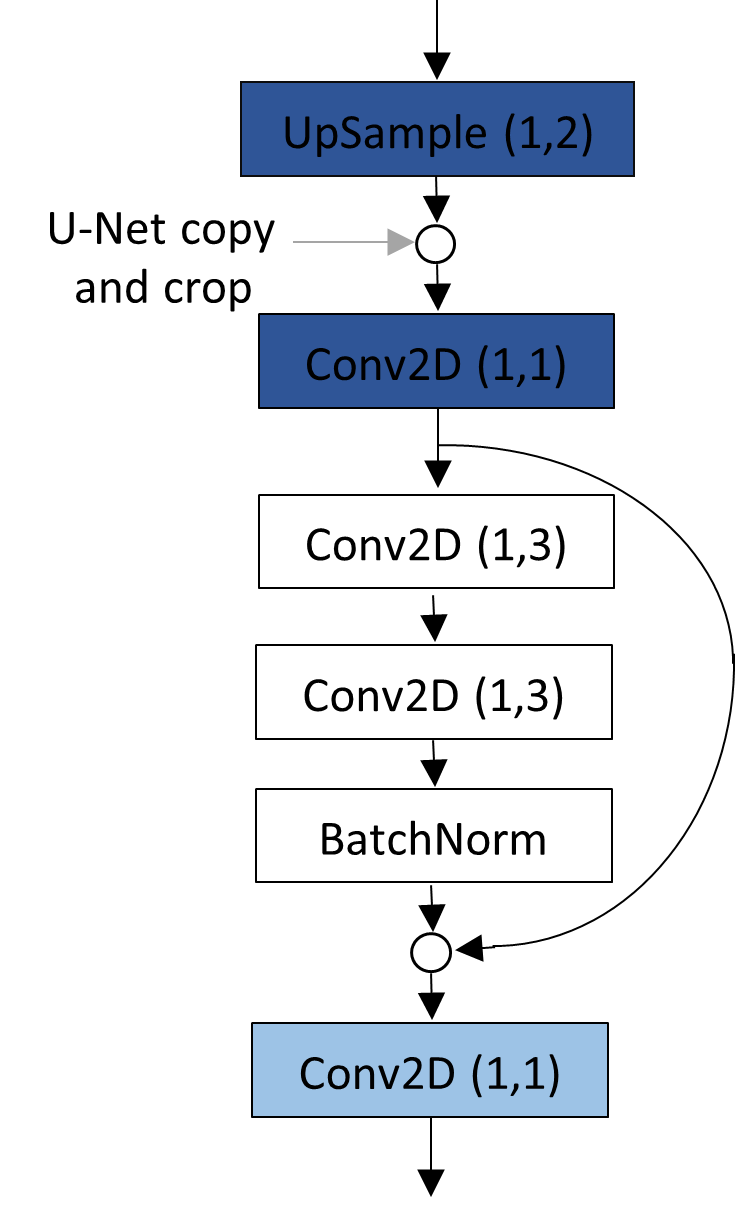}
     
            \label{expansion}}
		\end{subfigure}

\caption{Modified blocks}
\label{u-net block}
\end{figure}
    For dense prediction, we used the U-Net structure inspired by \cite{unet2015} and \cite{UNet2019} with modifications for our purpose. Similar to \cite{UNet2019} the input motion data is formatted like a one-column image with the channel size equal to the number of axis in motion data, and the upsampling is adopted instead of the convolution transpose. Different from previous work, we added batch normalization and skip connection followed by the convolution with a 1*1 kernel. The details are in figure \ref{u-net block}. The skip connection and 1*1 kernel are to weight and pool the features so the model could have access to both high-frequency and low-frequency features from the previous block and current block. To reduce the computation due to concatenation, the number of filters in the 1*1 kernel is the same as in previous convolution layers so the dimension is maintained. The final output is the convolution layers with a 1*1 kernel.

\subsubsection{Discriminator}

The discriminator takes the motion data and the dense labels from ground truth or dense predictions. For the discriminator, the input motion data is convoluted with (1,1) kernel to have the same dimension as the dense labels. Then, the motion data and the dense label pairs are passed to three blocks of the contraction block in Figure \ref{contraction} with kernel size (2,3). The output layer is convoluted with kernel size (1,1) to output patches of predictions.

        \subsubsection{Loss}
        
        Simply combining the methods may not be suitable for our application, so we customized the loss function adaptive for activity classification.
        Different from the categorical cross-entropy or $\mathcal{L}_1 $ loss in \cite{UNet2019,pix2pix2017}, we combined the focal loss with the multiplier$\beta $ and GAN loss, see equation \ref{ltotalGAN}.  $\lambda$ is a factor to balance the importance between $\mathcal{L}_{cGAN}$ and $\mathcal{L}_{focal} $. The focal loss is mainly to address skewness in the class distribution.

\begin{equation}\label{ltotalGAN}
            G^*=arg\:\displaystyle \min_{G}\:\max_{D}\beta\mathcal{L}_{cGAN}(G,D)+\lambda \mathcal{L}_{focal}(G)
\end{equation}
        During the adversarial training, the process does not always compete for better results when the focal loss stables and the loss for GAN dominate. Then, the discriminator could push the generator to generate dense labels with few fragmentations but more miss-classified instances.
        To adjust the amount of feedback from the discriminator, we added a discount value to the cGAN loss based on the correctly predicted location in the dense labels. When most of the dense prediction is correct, the feedback from the discriminator will be minimized. We used the dice coefficient between generated dense labels $ G(x)$and the ground truth $y$ in the equation \ref{multipler}. 
        \begin{equation}\label{multipler}
    \beta:\: 0<\mathcal{L}_{dice}=1-\frac{2|y\cup G(x)|}{|y|+|G(x)|}\leq 1
\end{equation}
      


\section{Experimental Setup}
\begin{table*}[ht]
\caption{Labeled Data Composition. The Minority Activities here is activities that are ten times smaller than the most activities.}

\begin{center}
\begin{small}

\begin{tabular}{lllll}
\hline
Dataset &Sampling Rate &Motion Sensor &Majority Activities  &Minority Activities  \\

\hline
Cow& 50Hz&Accelerometer &Grazing,Ruminating,Lying Standing&Walking
\\\hline
Pig&10Hz& Accelerometer, Gyroscope &Eating, Lying, Walking,&  Drinking,Standing\\\hline
\multirow{2}{*}{HAPT}  
&\multirow{2}{*}{50Hz} &\multirow{2}{*}{Accelerometer, Gyroscope}   & Standing, Sitting, Lying ,Walking, &Stand-to-sit,Sit-to-stand, Sit-to-lie, 
 \\&&&Walking downstairs ,Walking upstairs,& Lie-to-sit, Stand-to-lie,  Lie-to-stand
\\

\hline
\label{composition}
\end{tabular}
\label{selected}

\end{small}
\end{center}
\end{table*}
\begin{figure*}[!htb]
\centering
 \begin{subfigure}[Pig Dataset.]
		{\includegraphics[width=0.366\textwidth]{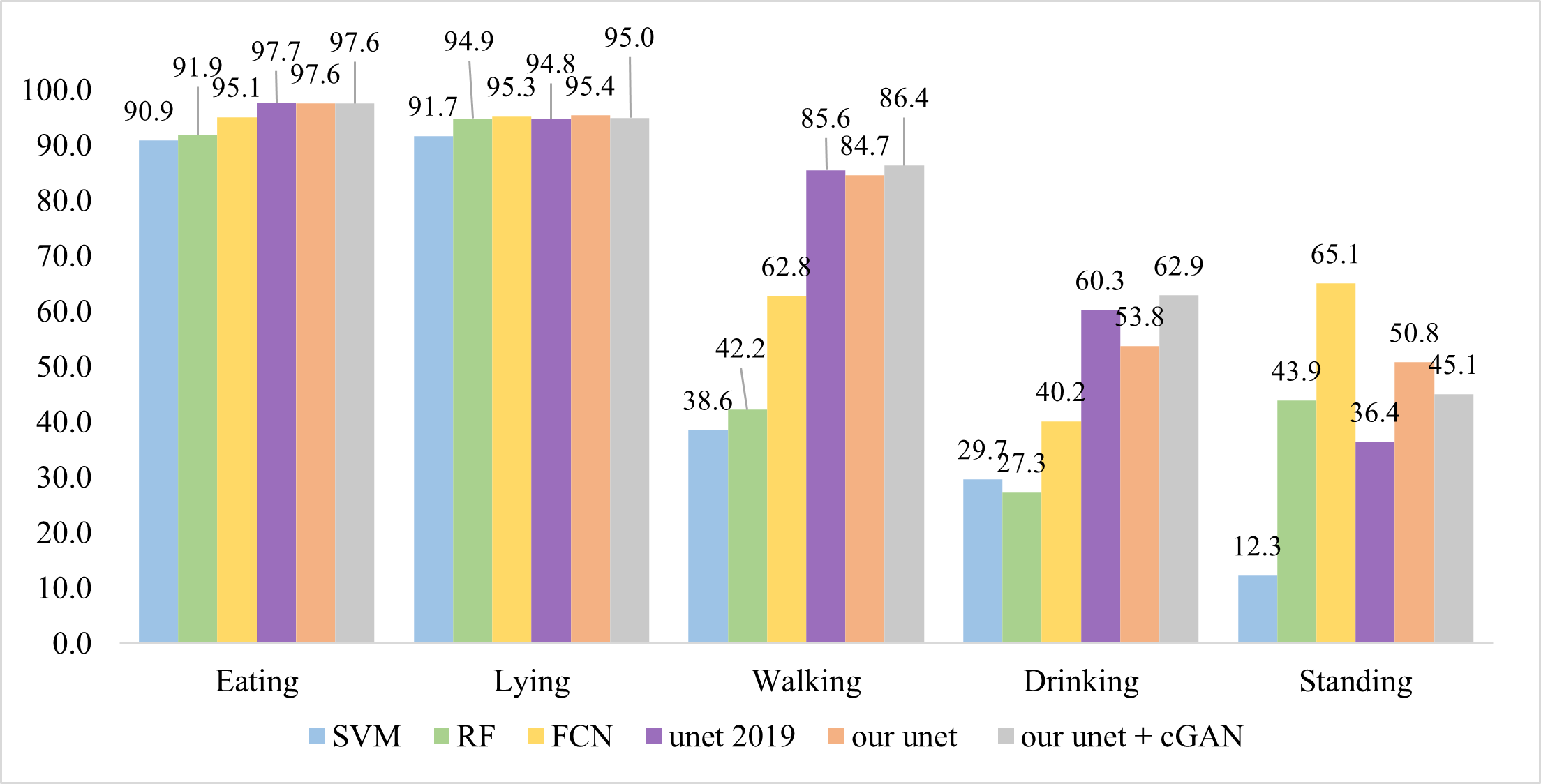}

            \label{contraction}}
		\end{subfigure}
	\begin{subfigure}[Cow Dataset.]
	{\includegraphics[width=.59\textwidth]{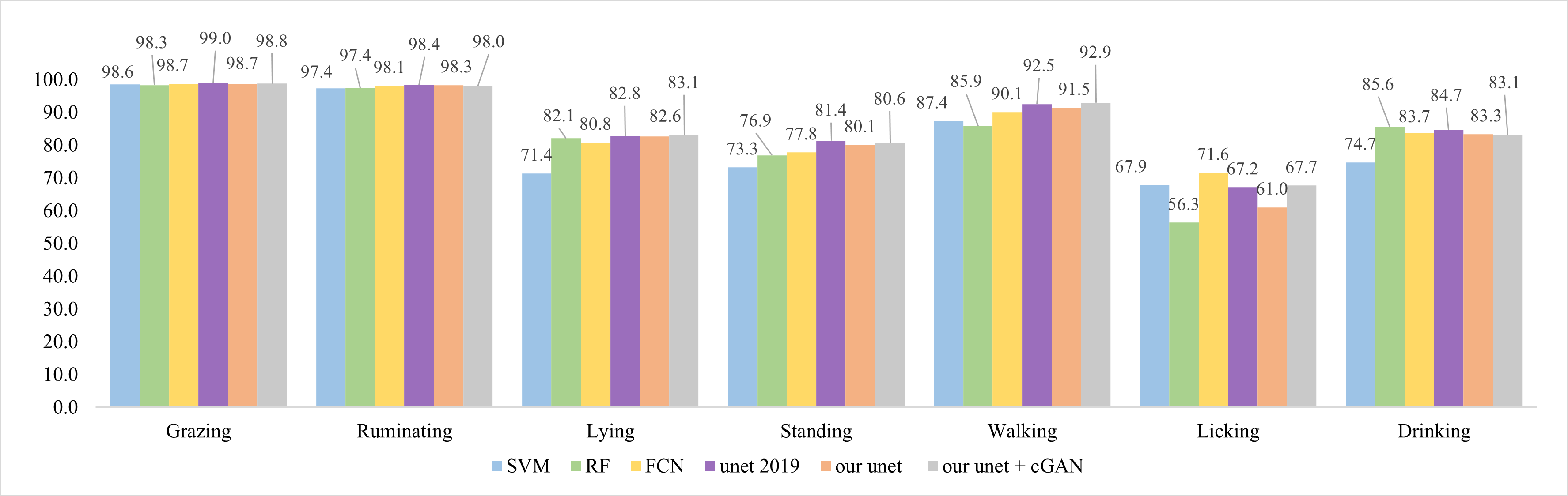}
     
            \label{expansion}}
		\end{subfigure}
    \begin{subfigure}[UCI HAPT Dataset.]
	{\includegraphics[width=.97\textwidth]{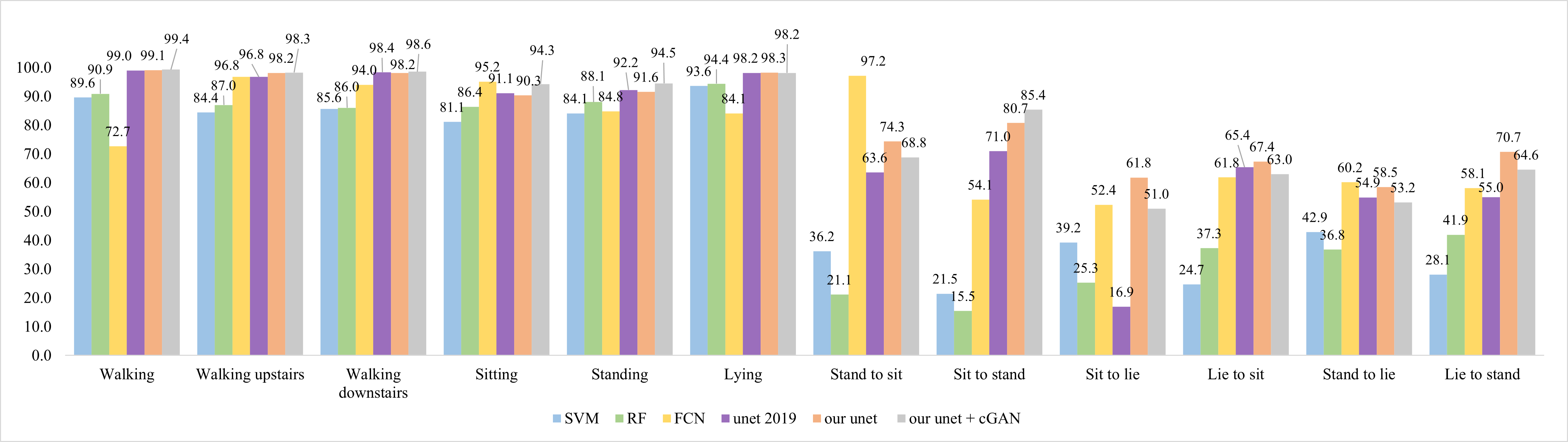}
     
            \label{expansion}}
		\end{subfigure}
\caption{The F1-scores on different classes of the methods on three datasets.}
\label{F1 score on individual class}
\end{figure*}

    \subsection{Dataset}

We test the models in our self-collected cow dataset, and pig dataset. For comparison, we test the models on the UCI HAPT dataset as well. The details of the pig, cow and HAPT dataset are in table \ref{composition}.

\subsubsection{Pig:} The pig dataset is \cite{salehpaper}. The WSN has two sensors, a 3-axis accelerometer, and a 3-axis gyroscope, with a 10 Hz sampling frequency to measure acceleration and angular velocity. The sensors are attached to the back using harnesses. 

\subsubsection{Cow:}The WSN is attached to the side of the halter. The WSN consists of a 3-axis accelerometer to collect motion data. We chose the 50 Hz sampling rate and a low-pass filter with a cutoff frequency of 24 Hz to reduce aliasing. The measurement of the accelerometer ranges from $-2g$ to $+2g$, where $1g$ is the gravity acceleration. We collected data on four different days, and on 5 female beef cows. The data collected on different days and different cows are normalized and concatenated to form the final dataset. 

\subsubsection{HAPT:} To compare the performance with the existing dense prediction works \cite{FCN2017,UNet2019}, we test our model on one of the same datasets UCI HAPT dataset \cite{HAPT} because the HAPT dataset contains stationary, active, and transitional activity with various activity duration and imbalanced distribution. 

    \subsection{Implementation Details}
    For all the three datasets, the data collected on the different objects and different days are normalized and concatenated to form the final dataset. Then, the data were randomly split into 43.56\% training, 21.78\% validation, and 34.66\% testing with random seed 42. The trained models are stored at the checkpoint when the performance is maximum on the validation set. The final comparison is on the test set.

    \subsubsection{Feature Extraction-Based Support Vector Machine (SVM) and Random Forest (RF)}
    The features are implemented based on the work \cite{ACM2018} including max, min, mean, standard deviation, median, $25^{th}$ percentile, $75^{th}$ percentile, mean low pass filtered signal, mean rectified high pass filtered signal, Skewness of the signal, Kurtosis, zero crossing rate, principal frequency, spectral energy, frequency entropy, and frequency magnitudes. The SVM is using the Radial Basis Function kernel and the RF is using 300 estimators with a maximum depth of 25 for each tree. The data is segmented into 256 window sizes and the corresponding labels are based on the most frequent activity. Then, the predicted labels are extended to the same window size and compared against ground truth dense labels.
    
    \subsubsection{FCN 2018}
    The FCN \cite{FCN2017} is implemented with six blocks of convolution followed by max pooling, a dropout layer with a drop-out rate of 0.2. We changed the output layer to a convolution transpose layer instead of the deconvolution layer. The input size is (1, 512, channels) for the cow and HAPT dataset.

     \subsubsection{U-Net 2019}
    For performance comparison, we used the code from U-Net \cite{UNet2019} and the same setting in the original paper: 32 filters, a drop-out rate of 0.2, and input size (1, 224, channels) for cow and HAPT dataset. The optimizer is Adam with categorical cross-entropy. The model is trained with reduce learning rate on plateau and 100 epoch. 
    \subsubsection{Modified U-Net}
    We used Parametric ReLU (PReLU) as activation \cite{prelu}. We used Adam optimizer with learning rate scheduler: exponential decay with 0.0005 initial rate, 0.96 decay rate, and 300k decay steps. There are 70k training steps, each batch contains 100 training instances. We added a spatial dropout with a drop-out rate of 0.2 in block 2 after skip connection for the HAPT dataset, and block 6 for the cow dataset. The input size is (1, 256, channels) for the cow and HAPT dataset. 
   
    All three models used the input size (1, 128, channels) for the pig dataset, because of the data scarcity. 

\section{Experimental Results}
We compare the performance of feature extraction-based SVM and RF, and dense prediction-based FCN, U-Net, U-Net trained with cGAN, modified U-Net, and modified U-Net trained with cGAN. In table \ref{overall performance}, the models trained with cGAN show improvements in the pig and UCI HAPT dataset. The U-Net accuracy improve from 92.23\% to 94.34\% and our modified U-Net accuracy improve from 93.66\% to 94.66\% for the HAPT dataset. The U-Net accuracy increased from 90.54\% to 93.08\% and our modified U-Net accuracy increased from 92.63\% to 93.18\% for pig data. The performance of cow data did not show significant improvements. The cow data may lack information to further distinguish between different activities, so the discriminator in cGAN did not provide sufficient feedback for generators. F1 score on individual class is in Figure \ref{F1 score on individual class}. The dense prediction methods are generally better than the feature extraction-based methods, especially for activities with shorter duration. In the HAPT and pig datasets, our modified model and model trained with cGAN generally perform well in the minority class: walking, drinking, sit to standing, sit to lie and lie to stand.
\begin{table}[!htb]
\centering

\begin{tabular}{l|llll}
\hline
&&Pig&Cow&HAPT\\
\hline
{\scriptsize \multirow{2}{*}{SVM}} &Acc&82.50&87.85&82.15\\&Fw&52.63&81.49&59.25
\\\hline
{\scriptsize \multirow{2}{*}{RF}} &Acc&85.65&90.66&84.56\\&Fw&60.03&83.21&59.22
\\\hline
{\scriptsize \multirow{2}{*}{FCN 2018}} &Acc&88.47&90.98&89.48\\&Fw&59.97&85.85&75.95
\\\hline
{\scriptsize\multirow{2}{*}{U-Net 2019} }&Acc&90.85&92.17&92.23\\&Fw&65.74&86.728&75.21
\\\hline
{\scriptsize cGAN +} &Acc&93.08&92.48&94.34
\\{\scriptsize  U-Net 2019}&Fw&75.65&86.81&80.70
\\
\hline

{\scriptsize{\scriptsize\multirow{2}{*}{Modified U-Net}}} &Acc&92.63&91.17&93.66\\ &Fw&74.35&86.02&78.69
\\\hline
{\scriptsize cGAN +} &Acc&93.18&91.89&94.66\\ {\scriptsize Modified U-Net}&Fw&77.38&86.31&81.15

\\\hline
\end{tabular}
\caption{The performance of models trained with and without cGAN in different dataset.}
\label{overall performance}
\end{table}
\subsection{Misalignments Analysis}
\begin{figure}[!htb]
\centering
\includegraphics[width=\columnwidth]{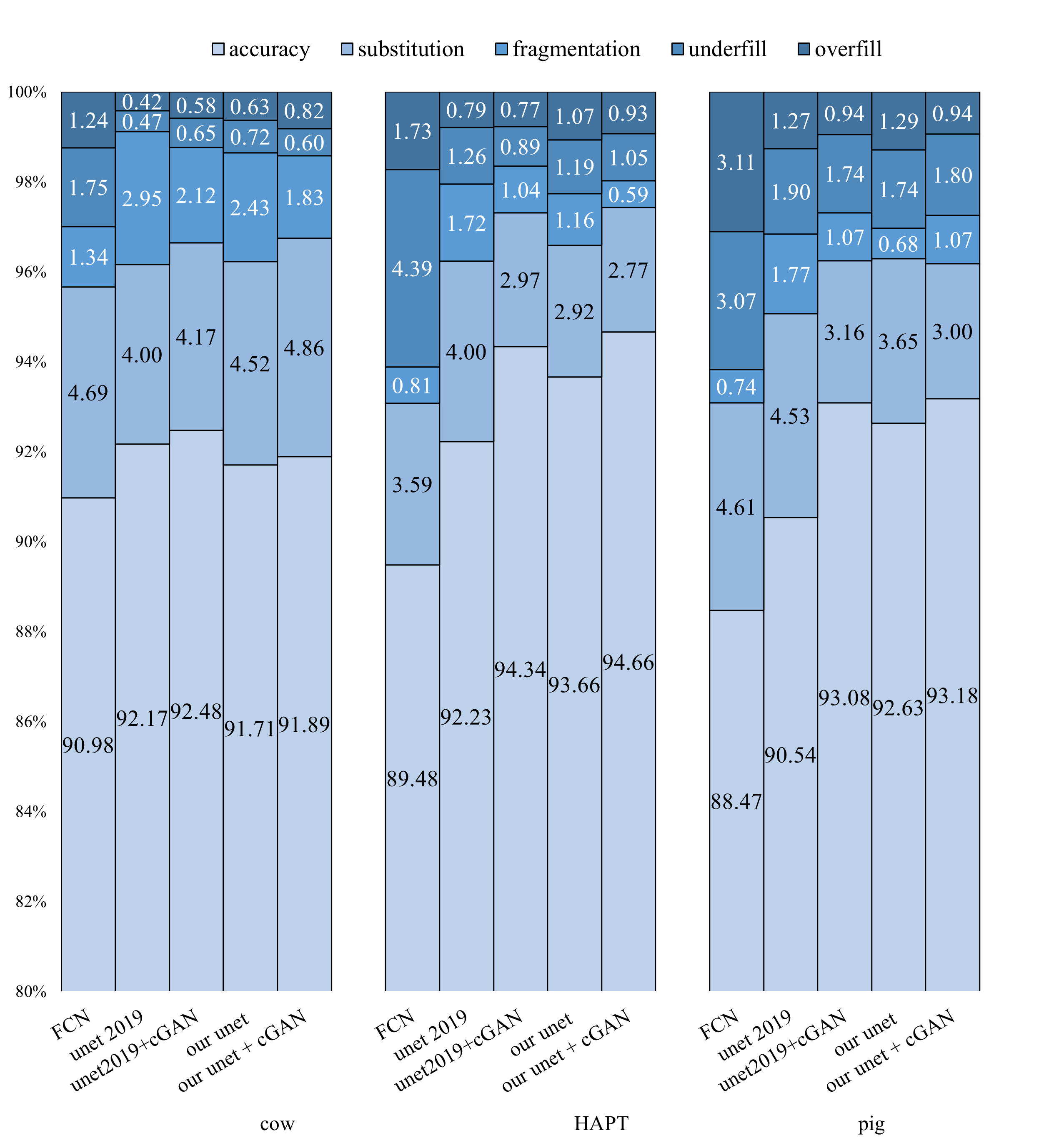} 
\caption{The composition of four types of misalignments}
\label{error composition}
\end{figure} 
To further analyze the results, we plot the composition of misalignments in figure \ref{error composition}. In the three datasets, most of the fragmentation rate decreased even for the cow data: from 2.95\% to 2.12\% and from 2.43\% to 1.83\%. This decrease would mostly result in the cGAN structure. The overfill and underfill rate slightly decreased in HAPT and pig dataset, but increased in the cow dataset. The substitution rate decreased from 4.00\% to 2.97\%, 2.92\% to 2.77\% in HAPT dataset and from 4.53\% to 3.16\%, 3.65\% to 3.00\% in pig dataset. Summarily, 
the model trained with cGAN generally tries to force the generator to predict dense labels with smaller fragmentation errors, when the discriminator could provide discriminative power. The resulting models could achieve higher performance. When there is not enough information such as the cow dataset, model trained with or without cGAN tends to have similar overall performance with a decrease in fragmentation rate but an increase in other misalignments error. 

\section{Conclusion}
We modified and experimented with dense labeling and prediction for animal activity classification purposes. The dense prediction model could provide activity durations and indicate the time location of activities. Existing works may not work well for activities with longer durations because the inter-similarity between activities and the exact starting and ending points are hard to define. We proposed a cGAN to train the generator, which is the modified U-Net, so the discriminator could push the generator to reduce fragmentation and other errors. To prevent the discriminator from penalizing the generator too much and increasing the over and underfill rate, we used the dice coefficient as a discount value to the cGAN loss. When the dense prediction map reaches the optimum performance, the feedback from the discriminator is minimized. After testing with the pig, cow, and HAPT datasets, the model trained with cGAN general achieves higher or comparable performance, especially for the activities with short duration and imbalanced distribution. 

The future work would be effectively extracting information from different sensors so the cGAN method could be beneficial, especially in reducing the substitution rate. Then, we could apply the dense prediction method to automatic labeling.


\bibliography{aaai22}

\end{document}